\documentclass[aps,showpacs,twocolumn,prl,superscriptaddress]{revtex4}
\usepackage{graphicx}
\input amssym.tex

\begin{document}
\title{Observation of a Chiral State in a
Microwave Cavity}
\author{C.~Dembowski}
\affiliation{Institut f{\"u}r Kernphysik, Technische Universit{\"a}t
Darmstadt, D-64289 Darmstadt, Germany}
\author{B.~Dietz}
\affiliation{Institut f{\"u}r Kernphysik, Technische Universit{\"a}t
Darmstadt, D-64289 Darmstadt, Germany}
\author{H.-D.~Gr{\"a}f}
\affiliation{Institut f{\"u}r Kernphysik, Technische Universit{\"a}t
Darmstadt, D-64289 Darmstadt, Germany}
\author{H.L.~Harney}
\affiliation{Max-Planck-Institut f{\"u}r Kernphysik, D-69029
Heidelberg, Germany}
\author{A.~Heine}
\affiliation{Institut f{\"u}r Kernphysik, Technische Universit{\"a}t
Darmstadt, D-64289 Darmstadt, Germany}
\author{W.~D.~Heiss}
\affiliation{Department of Physics, University of Stellenbosch, 7602
Matieland, South Africa}
\author{A.~Richter}
\email{richter@ikp.tu-darmstadt.de} \affiliation{Institut f{\"u}r
Kernphysik, Technische Universit{\"a}t Darmstadt, D-64289 Darmstadt,
Germany}
\date{\today}

\begin{abstract}
A microwave experiment has been realized to measure the phase
difference of the oscillating electric field at two points inside
the cavity. The technique has been applied to a dissipative
resonator which exhibits a singularity -- called exceptional point
-- in its eigenvalue and eigenvector spectrum. At the singularity,
two modes coalesce with a phase difference of $\pi/2\, .$ We
conclude that the state excited at the singularity has a definitiv
chirality.

\end{abstract}
\pacs{05.45.Mt, 41.20.Jb, 03.65.Vf, 02.30.-f}
\maketitle

Recently a surprising phenomenon occurring in systems described by
non-hermitian Hamiltonians has received considerable attention: the
coalescence of two eigenmodes. If the system depends on some
interaction parameter $\lambda\, ,$ the value $\lambda_{\rm EP}$ at
which the coalescence occurs is called an exceptional point (EP)
\cite{Kato}. At an EP, the eigenvalues \emph{and} eigenvectors show
branch point singularities \cite{Kato,Hesa,BerryEP,Mondragon,Rotter}
as functions of $\lambda\, .$ This stands in sharp contrast to
two-fold degeneracies, where no singularity but rather a diabolic
point \cite{BerryDP} occurs. EPs have been observed in laser induced
ionization of atoms \cite{Latinne}, acoustical systems \cite{Shuva},
microwave cavities \cite{BrentanoEP,DemboEP}, certain absorptive
media \cite{Pancha}, and in ``crystals of light'' \cite{Oberthaler}.
The broad variety of systems showing EPs indicates that their
occurrence is \emph{generic} which is discussed further in
\cite{BerryEP,BerryPancha}. So far EPs have been observed in decaying
systems described by a complex symmetric effective Hamiltonian
\cite{WW}. While the theoretical and experimental articles cited
above discuss the properties of systems in the vicinity of an EP, a
recent theoretical work \cite{PhaseTheo} investigates the complex
symmetric Hamiltonian of a two-level system \emph{at} the EP. The
eigenfunction at the EP turns out to be
\begin{equation}
|\psi_{\rm EP}\rangle \propto |1\rangle \pm i|2\rangle
\label{PsiEP}
\end{equation}
for \emph{any} choice of the basis states $|1\rangle$ and
$|2\rangle\, .$ This is a chiral state: in quantum, acoustical and
electromagnetic systems the two orthogonal basis states oscillate
in time; if they are superimposed according to Eq.~(\ref{PsiEP})
-- where they follow each other with a time lag of a quarter
period -- the result is rotating either clockwise or
counter-clockwise. This is in analogy with the generation of
circularly polarized light being a superposition of two orthogonal
linearly polarized waves phase shifted by $\pi/2$. The sign of the
chirality is defined via the direction of time and in the
experiment the positive direction of time is given by the decay of
the eigenstates. The first observation of such a chiral state in a
microwave cavity experiment is the gist of the present paper.

\begin{figure}[ht]
\includegraphics[width=7.5cm]{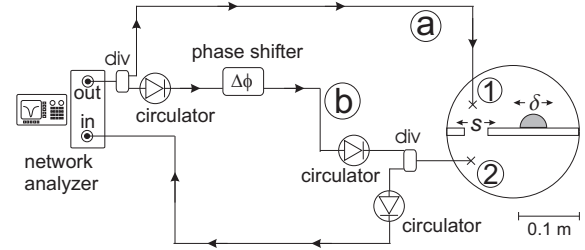}
\caption{Experimental setup to measure the phase shift between two
different positions in the cavity. The geometry of the resonator can
be changed by adjusting the widths $s$ of a slit between the two
halves (labeled (1) and (2), resp.) and the position $\delta$ of a
teflon stub. Microwave power is coupled into the resonator via path
(a) and antenna 1 into part (1) and, with a tunable phase shift
$\Delta \phi$, via  antenna 2 into part (2), where it is also picked
up.} \label{Expsetup}
\end{figure}

The high precision of such an experiment makes it a prime choice to
observe the chirality of $|\psi_{\rm EP}\rangle$, yet measuring a
phase difference of $\pm \pi/2$ between two superimposed modes is an
unusual and demanding experimental task. A setup had to be designed
that allows to excite the two modes with an adjustable frequency
difference $f_1-f_2\, ,$ and in the case of $f_1=f_2$ with an
adjustable \emph{phase difference} between them. The cavity is
composed of two almost identical semi-circular parts, see Fig.
\ref{Expsetup}. They are coupled by a slit of variable width  $s\, .$
In order to reach $f_1=f_2$ one needs a second tunable parameter,
namely the position $\delta$ of a teflon semi-circular stub in part
$1$ of the cavity on  Fig. \ref{Expsetup}.
These two parameters make
up the interaction parameter $\lambda(s, \delta )\, .$ The two
eigenmodes coalesce at a critical value $\lambda_{\rm EP} =(s^{\rm
EP}, \delta^{\rm EP})$. In the sequel we chose $\lambda$ to be close
enough to $\lambda_{\rm EP}$ so as to consider the cavity a
two-state-system. This is possible when the distance between the two
coalescing eigenvalues is much smaller than the distance to any third
one. As in \cite{DemboEP}, the EP has been found by looking at the
behavior of the real and imaginary parts of the eigenvalues, i.e. the
resonance frequencies $f_1,\, f_2$ and widths $\Gamma_1,\, \Gamma_2\,
.$ For weak coupling, i.e. $s < s^{\rm EP}$, one observes a crossing
of $f_1,\, f_2$ and an avoided crossing of $\Gamma_1,\, \Gamma_2$
when one sweeps $\delta\, .$ For strong coupling, i.e. $s > s^{\rm
EP}\, ,$ an avoided crossing of $f_1,\, f_2$ and a crossing of
$\Gamma_1\, ,\Gamma_2$ takes place \cite{Hesa,DemboEP,he99}. It is
only \emph{at} the EP that one expects the real \emph{and} imaginary
parts of the eigenvalues to be equal. In addition to the eigenvalues,
the eigenfunctions were studied by mapping the distributions of the
electric field. When the EP is encircled in the space of $\lambda\,
,$ \emph{one} of the eigenvectors undergoes a change of sign
\cite{DemboEP}, i.e. it picks up a so-called ''geometric phase''
\cite{BerryBuch}. Which eigenvector changes its sign is defined by
the only freedom left: the orientation of the closed loop around the
EP. This is a first fingerprint of the definite chirality of the EP.

\begin{figure}[ht]
\includegraphics[width=7.5cm]{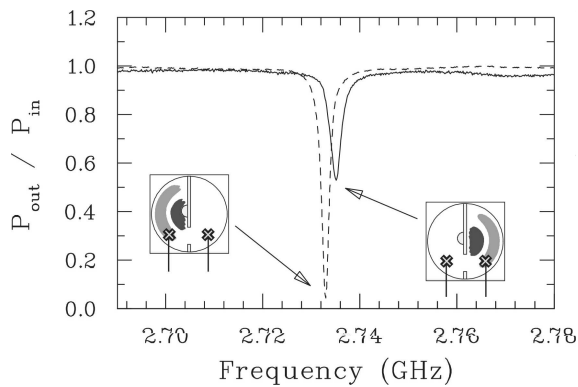}
\caption{Reflection spectra, i.e. the outcoupled power over the
incoupled one as a function of frequency, taken with two different
antennas in mirrored positions marked by crosses with attached lines.
The full (dashed) line is taken with the antenna in the right (left)
semicircular part of the cavity. In the frequency interval shown lie
two eigenmodes of the resonator with field distributions plotted as
insets, however, in each spectrum only \emph{a single} resonance is
visible since each antenna couples to only one of the two
eigenmodes.} \label{Modes}
\end{figure}

For $s < s^{\rm EP}$ the two modes discussed in the following are
each \emph{localized} in one of the semi-circular parts of the cavity
--  as the measured field distributions show (cf. insets of
Fig.~\ref{Modes}). We consider these localized modes as the basis
configurations $|1\rangle$ and $|2\rangle$ of the two-state system.
They can be excited separately even if $f_1 = f_2$ by appropriately
positioned dipole antennas \cite{RichterBuch} which are located at
mirror positions in the cavity. The coupling between the antennas and
the resonator is proportional to the square of the electric field at
the locations of the antennas \cite{RichterBuch,StoeckmannPhase}. If
the electric field is approximately zero at the location of a given
antenna the mode cannot be excited. In Fig.~\ref{Modes} the
reflection spectra for two different antennas are shown for $s=38\,
$mm. The parameter $\delta$ is tuned so that that $f_1$ is slightly
different from $f_2\, .$ Still each spectrum shows only a single
resonance -- not a doublet. It follows that each antenna indeed
excites only one basis configuration.

In the two-state-regime, the eigenstates $|\psi_k\rangle$ of the
cavity are expanded according to
\begin{equation}
|\psi_k\rangle =a_{k1}|1\rangle + a_{k2}|2\rangle\, ,
\label{eigenst}
\end{equation}
however, for sufficiently small $s$, the eigenmodes
$|\psi_k\rangle$ approach the localized modes $|1\rangle\, ,$ and
$|2\rangle\, $ as can be seen in Fig.~\ref{Modes}. The relative
phase of the amplitudes $a_{k1}$ and $a_{k2}$ of $|\psi_1\rangle$
and $|\psi_2\rangle$ has been measured by the following technique,
applied -- to the best of our knowledge -- for the first time in
microwave cavities. A microwave source, an HP8510C vectorial
network analyzer is emitting continuous rf-radiation with a fixed
frequency, split by a power divider (Narda 4313-2) and fed into
the cavity via two paths labeled (a) and (b) in
Fig.~\ref{Expsetup}. The phase of the signal travelling through
(b) can be shifted by $\Delta\phi$ using a Narda 3752
phaseshifter. Several circulators \cite{Circulator} are used to
suppress reflections along path (b). The amplitude $S$ of the
signal outcoupled by antenna 2 is a superposition of two coherent
waves with different phases, i.e.
\begin{equation}
S \propto t + r e^{i\Delta\phi}\, .
\label{Signal}
\end{equation}
Here, $t$ and $r$ denote the transmission and reflection coefficients
of the cavity. For further analysis, two cases have to be
distinguished: First, for $s<s^{\rm EP}$, $\delta$ is tuned in such a
way that the frequencies of the two eigenstates $|\psi_1\rangle$ and
$|\psi_2\rangle$ are the same and both are excited in resonance by
the two signals coupled to the cavity. Second, for $s=s^{\rm EP}\, ,$
the \emph{only} state present, $|\psi_{\rm EP}\rangle$, is excited by
both signals.

In both cases, $t$ and $r$ are known from scattering theory (see
e.g. \cite{SMatrix1,SMatrix2,SMatrix3}) and the dependence of the
signal received by the network analyzer from $\Delta\phi$ is given
by
\begin{equation}
|S(\Delta\phi)|^2 \propto C+\cos^2\left(\frac{\Delta\phi -
\phi_0 - \tilde{\phi}}{2} \right)\, .
\label{RefPower}
\end{equation}
Here, the constant $C\ge 0$ determines the contrast of the
pattern. When $C$ is large, the contrast is small, which occurs
whenever the basis states contribute very unevenly to the
eigenmodes. The angle $\phi_0$, i.e. the phase difference between
the oscillating fields at the position of the antennas, is given
by the relative phases of the expansion coefficients in
(\ref{eigenst}), i.e.
\begin{equation}
\phi_0 = \arg (a_{12}/a_{11})
\label{phasey1}
\end{equation}
for an isolated eigenmode and
\begin{equation}
\phi_0 = \arg \left(\frac{a_{12}+a_{22}}{a_{21}}\right)
\label{phasey}
\end{equation}
for the superposition of two eigenmodes. If Eq.~(\ref{PsiEP})
holds true we expect $\phi_0 = \pm \pi/2$ for $|\psi_{\rm
EP}\rangle$, proving the chirality of $|\psi_{\rm EP}\rangle$.

\begin{figure}[ht]
\includegraphics[width=7.5cm]{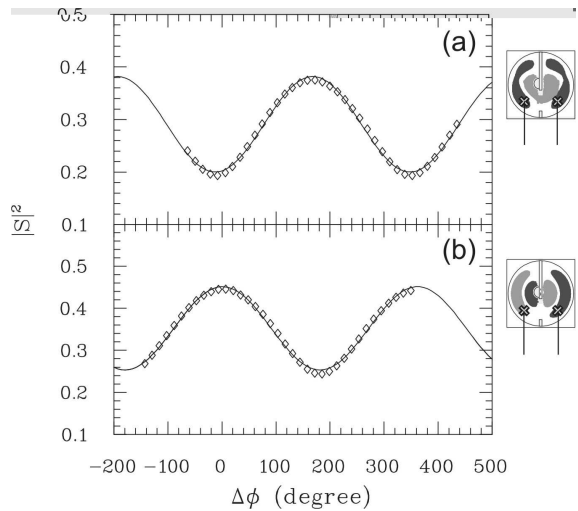}
\caption{The square of the amplitude of the signal outcoupled by
antenna 2, $|S|^2$, as a function of the phase shift $\Delta\phi$. In
(a) the phase difference $\phi_0$ of the oscillating electric field
between the antennas (marked as crosses with attached lines) is $0$
for the measured field distribution shown, while in (b) it is $\pi$.
In both cases the minimum of $|S(\Delta\phi)|^2$ corresponds to
$\phi_0$. } \label{TestSetup}
\end{figure}

The angle $\tilde{\phi}$ in Eq.~(\ref{RefPower}) is due to the
remaining length difference of the paths (a) and (b) and the fact
that the antennas (1) and (2) may cause additional and different
phase shifts. It has to be determined by analyzing an eigenstate with
known $\phi_0\, .$ One easily finds \cite{StoeckmannPhase} modes with
$\phi_0 =0$ or $\pi\, .$ Figure \ref{TestSetup}a shows
$|S(\Delta\phi)|^2$ for a mode with $\phi_0=0$, i.e. the electric
field at the position of the two antennas oscillates \emph{in phase},
cf. Fig.~\ref{TestSetup}a, where the antennas are sketched as crosses
with attached lines.
By fitting Eq.~(\ref{RefPower}) to the measured
$|S(\Delta \phi)|^2$ we find $\tilde{\phi}\approx \pi$, which is
expected from measurements of the reflection coefficients
\cite{padamsee} of the antennas which shift the phase of an emitted
wave by $\pi$. In Fig.~\ref{TestSetup}b we show $|S(\Delta\phi)|^2$
for a mode with $\phi_0=\pi$, as can be seen from the measured field
distribution. The data agrees again well with Eq.~(\ref{RefPower}).

\begin{figure}[ht]
\includegraphics[width=7.5cm]{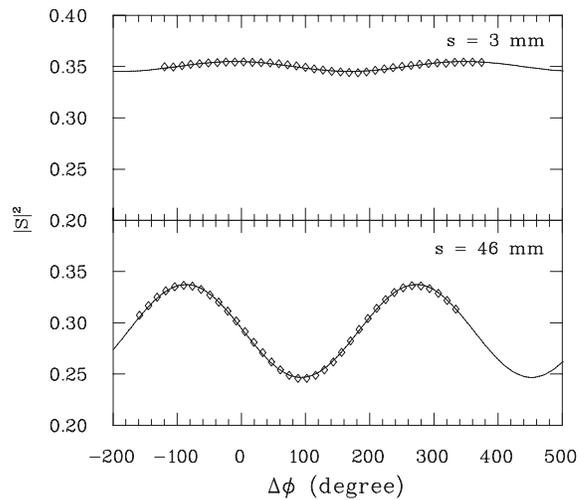}
\caption{The square of the amplitude of the signal outcoupled by
antenna 2, $|S|^2$, as a function of the phase shift $\Delta\phi$ for
weak coupling $(s=3\, {\rm mm})$ and at the EP $(s=46\, {\rm mm})$.
As in Fig.~\ref{TestSetup} the minimum of $|S(\Delta\phi)|^2$
corresponds to the phase difference of the oscillating electric field
between the antennas.} \label{Contrast}
\end{figure}

Having verified that this setup allows to measure the phase shift
between different points of a field distribution, we turn to the
investigation of $|\psi_{\rm EP}\rangle$. To reach the EP the slit is
at first set to $s = 3\, {\rm mm}$, i.e. $s < s^{\rm EP}\, .$ The
teflon semi-circle can be moved from the outside, i.e. \emph{while}
microwave power is coupled into the cavity and therefore $\delta$ can
be tuned such that $f_1 = f_2\, .$ Then the pattern $|S(\Delta\phi
)|^2$ exhibits only a weak contrast (see the upper part of Fig.
\ref{Contrast}).
This occurs because $|a_{12}|$ is very small compared to $|a_{11}|$
and $|a_{22}|$ is large compared to $|a_{21}|$ for such a weak
coupling. Nevertheless, $\phi_0$ can be obtained by fitting
Eq.~(\ref{RefPower}) to the measured data. It is close to $\pi$ for
$s<28\, {\rm mm}$, cf. Fig.~\ref{PhaseMeasurement}, which agrees with
general arguments given below.

Secondly we set $\lambda \approx \lambda_{\rm EP}$ so that
$|\psi_{\rm EP}\rangle$ is the only mode present (lower part of
Fig.~\ref{Contrast}). The contrast of the pattern $|S(\Delta\phi)|^2$
is high because the expansion coefficients of $|\psi_{\rm EP}\rangle$
have the same absolute value. For $s=46\, {\rm mm}$ one obtains
$\phi_0 =\pi/2\pm 0.05\, ,$ cf. Fig. \ref{PhaseMeasurement}. This is
the phase predicted in Eq.~(\ref{PsiEP}) -- constituting the first
experimental observation of a mode, were the electric field between
two points oscillates with a phase difference of $\pi/2$. As pointed
out, this is a chiral mode rotating clockwise in the basis frame
spanned by $|1\rangle$ and $|2\rangle$.

Figure \ref{PhaseMeasurement} shows the behavior of $\phi_0$ from
small to large slit openings. For each value of $s\, ,$
$|S(\Delta\phi)|^2$ has been taken as displayed on Fig.
\ref{Contrast} by sweeping $\Delta\phi$ over $\approx 400^{\circ}\,
.$ The error of $\phi_0$ is a systematic error caused by varying
bending radii of the coaxial microwave cables and the corresponding
variation of the length of the wave paths during the measurement. The
surface properties of the connections between the cables and the
antennas -- which cannot be exactly controlled -- are another source
of error.

For $43 \; {\rm mm} < s < 48 \; {\rm mm}$ a crossing of the resonance
frequencies \emph{and} widths is observed and we conclude that this
is the precision with which the present experiment yields $s^{\rm
EP}\, .$ This does not contradict the theory \cite{Hesa} -- the
experiment simply cannot resolve the tiny avoided frequency or widths
crossings that occur for $s$ very near to $s^{\rm EP}\, .$
\begin{figure}[ht]
\includegraphics[width=7.5cm]{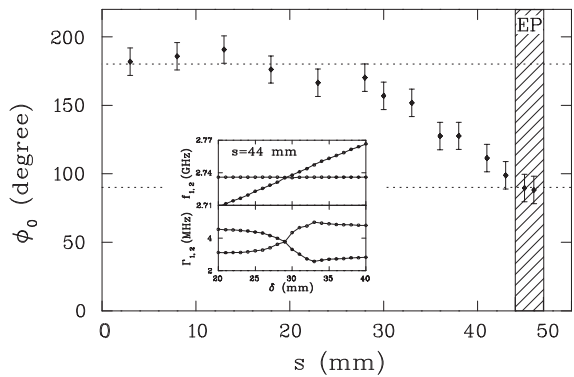}
\caption{The phase difference $\phi_0$ of the oscillating electric
field between the antennas for different couplings $3\, {\rm mm} \leq
s \leq s^{\rm EP}$. The EP is reached for $44\, {\rm mm} \leq s \leq
47\, {\rm mm}$ (shaded area), where we find, within the resolution of
the experiment, $f_1 \approx f_2$ and $\Gamma_1 \approx \Gamma_2$.
The phase difference at the EP is $\phi_0=(90\pm 3)^{\circ}$,
indicating the chirality of $|\psi_{\rm EP}\rangle$. As an inset we
show the crossing of the resonance frequencies \emph{and} widths for
$s=44\, {\rm mm}\approx s^{EP}$}
\label{PhaseMeasurement}
\end{figure}

For better understanding of the results of Figs. \ref{Contrast}
and \ref{PhaseMeasurement} let us recapitulate the two-state model
of \cite{PhaseTheo}. The Hamiltonian
\begin{equation}
H=\pmatrix{E_1-i\gamma_1 & H_{12}        \cr
           H_{12}        & E_2-i\gamma_2}
\label{ham}
\end{equation}
is complex symmetric. All its entries are functions of the
geometric parameters $\lambda\, .$ An EP occurs whenever the
eigenvalues of $H$ coincide and $H_{12}\ne 0\, .$ The sum of the
relative amplitudes occurring in Eq.~(\ref{phasey}) is
\begin{equation}
\frac{a_{12}}{a_{11}}+\frac{a_{22}}{a_{21}}
=\frac{\Delta E-i\Delta\gamma}{H_{12}}\, ,
\label{y}
\end{equation}
where $\Delta E = E_1-E_2$ and $\Delta\gamma =\gamma_1-\gamma_2\,
.$ At the EP, Eq.~(\ref{y}) becomes
\begin{equation}
\frac{a_{12}}{a_{11}}+\frac{a_{22}}{a_{21}}
=\pm 2i\, .
\label{yEP}
\end{equation}
The argument of the l.h.s of Eq.~(\ref{yEP}) coincides with the r.h.s
of Eq.~(\ref{phasey}) if $a_{11}=a_{21}$ as is the case in the
experiment. For a weak coupling the r.h.s. of Eq.~(\ref{y}) is large
because $H_{12}$ approaches zero linearly with $s\, .$ As we enforce
equality of the real parts of the eigenvalues of $H$, a particular
weak coupling limit is specified. It implies $\Delta E=0\, .$ The
finding $\phi_0\approx\pi$ in Fig. \ref{PhaseMeasurement} means that
$H_{12}$ tends towards zero along the imaginary axis. This describes
just the coupling within a purely absorptive system.

In summary, we have performed a microwave cavity experiment to
measure the phase difference $\phi_0$ between two points of a field
distribution inside the resonator. The technique has been applied to
the superposition of two eigenmodes on a path to an EP. At the EP,
the modes coalesce into the single, non-localized mode $|\psi_{\rm
EP}\rangle=|1\rangle + i |2\rangle\, ,$ i.e. $\phi_0=\pi/2\, .$ The
phase $\phi_0$ is determined from the pattern $|S(\Delta\phi)|^2$ of
interference between the amplitude for transmission through the
cavity and a wave with the arbitrary phase shift $\Delta\phi\, .$ To
our knowledge, it represents the first experimental observation of a
mode in a microwave resonator, where the electric field oscillates
with a phase difference of $90^\circ$ at two different points.

The decay of the eigenmodes is essential for the EP to occur.
Therefore the experiment has been performed at room temperature
rather than under conditions of superconductivity \cite{RichterBuch}.
The irreversibility of the decay causes $|\psi_{\rm EP}\rangle$ to
change under the operation of time reversal. This should not be
confused with fundamental time reversal symmetry breaking observed
e.g. in the system of the neutral Kaons \cite{CPLEAR}. There the
Hamiltonian of the decaying two-state system is not complex
symmetric. Similarly, the chirality of $|\psi_{\rm EP}\rangle$, i.e.
the fact that $|1\rangle + i|2\rangle$ is rotating clockwise, is not
due to a parity violating interaction as in $\beta$-decay \cite{Wu}.
It rather is a property of the EP that we have investigated. The set
of EPs that one can find in the resonator is expected to exhibit both
chiralities \cite{Hesa}. In this sense the present chirality closely
resembles the chirality of molecules \cite{Pasteur}, underlining the
importance and the generic aspects of EPs even further.

We are particularly  grateful to T.H. Seligman for discussing the
results with us. C.D., B.D., and A.H. thank CONACyT for financial
support during the workshop on {\it Chaos in few and many body
problems} at CIC. This work has been supported by the DFG under
contract number Ri 242/16-3.

\end{document}